\documentclass[prd,aps,twocolumn,showpacs,floats,floatfix]{revtex4}
\usepackage[dvips]{graphicx}
\usepackage{amssymb}

\begin{document}

\title{Radial oscillations and stability of compact stars in Eddington-inspired Born-Infeld gravity}

\author{Y.-H. Sham, L.-M. Lin, P.~T. Leung}
\affiliation{Department of Physics and Institute of Theoretical Physics,
The Chinese University of Hong Kong, Hong Kong SAR, China}

\date{\today}

\begin{abstract}
We study the hydrostatic equilibrium structure of compact stars in 
the Eddington-inspired Born-Infeld gravity recently proposed by 
Ba\~{n}ados and Ferreira~[Phys. Rev. Lett. {\bf 105}, 011101 (2010)]. 
We also develop a framework to study the radial perturbations and stability
of compact stars in this theory. 
We find that the standard results of stellar stability still hold in this 
theory. The frequency square of the fundamental oscillation mode vanishes 
for the maximum-mass stellar configuration. 
The dependence of the oscillation mode frequencies on the coupling parameter 
$\kappa$ of the theory is also investigated. We find that the fundamental mode 
is insensitive to the value of $\kappa$, while higher-order
modes depend more strongly on $\kappa$. 
\end{abstract}



\maketitle

\section{Introduction}
\label{sec:Intro}

General relativity (GR) has been the most successful and popular theory 
of gravity in the past century. From its classic predictions on the 
perihelion advance of Mercury and deflection of light, to the later
predictions such as the orbital decay of the Hulse-Taylor binary pulsar due 
to gravitational-wave damping, GR has passed the experimental observations 
in these weak-field situations with flying colors. 
Testing the predictions of GR in strong-field situations, 
such as the final stage of binary black hole coalescence, will come soon 
from the detection of gravitational waves (see~Ref. \cite{Will2006} for a 
review on experimental tests of GR). 

Despite the great success of GR, the idea that GR may not be the correct 
theory to describe the Universe on cosmological scales is also gaining 
attention recently. This is due to the fact that, if GR is 
correct, we must then require the Universe to be dominated by some 
unknown component, called dark energy, in order to explain the accelerating 
expansion of the Universe. In the past decade, various alternative theories 
of gravity which deviate from GR on cosmological scales have been proposed 
in order to explain cosmological observations (see Ref. \cite{Clifton2012} for a 
recent review).

On the other hand, it is also well known that GR is not complete because 
of its prediction of spacetime singularities in the big bang and those 
inside black holes. It is generally believed that a consistent theory of 
quantum gravity is needed to resolve this issue. Recently, a new 
Eddington-inspired Born-Infeld (EiBI) theory of gravity was proposed by 
Ba\~{n}ados and Ferreira \cite{Ferreira10}. 
The appealing properties of this theory are that it is equivalent to GR in 
vacuum and can avoid the big bang singularity \cite{Ferreira10}. 
The theory differs from GR only in the presence of matter, and 
in particular the deviation becomes significant at high densities. 
It is thus reasonable to expect that compact stars may be the best 
astrophysical laboratories to test EiBI gravity. 
In fact, Pani {\it et al.}~\cite{Pani11,Pani12} have recently studied the 
structure of compact stars in EiBI gravity. 

In this work, we report our investigation of compact stars in EiBI gravity. 
We first extend the work of Pani {\it et al.}~\cite{Pani11,Pani12} by studying 
the static structure of compact stars in this theory using a large set of 
realistic equations of state (EOS). 
Furthermore, we develop a framework to study the radial perturbations and 
stability of these stars in this theory for the first time.

The plan of the paper is as follows: In Sec. ~\ref{sec:EiBI} we briefly 
summarize EiBI gravity. In Sec.~\ref{sec:Static} we derive the equations for 
constructing static equilibrium stars in EiBI gravity. 
In Sec.~\ref{sec:perturbation} we present the linearized equations for 
radial oscillations of compact stars. 
Section ~\ref{sec:eos} presents the technique we use to employ 
realistic EOS models in this study. In Sec.~\ref{sec:results} we present our
numerical results. Finally, our conclusions are summarized in 
Sec.~\ref{sec:conclusion}. 
We use units where $G=c=1$ unless otherwise noted.

\section{Eddington-inspired Born-Infeld gravity}
\label{sec:EiBI}

Here we briefly summarize the EiBI gravity proposed by Ba\~{n}ados and 
Ferreira~\cite{Ferreira10}. 
The theory is based on the following action
\begin{equation}
S = {1 \over 16 \pi} {2 \over \kappa} \int d^4x \left( 
\sqrt{ \left| g_{\mu\nu} + \kappa R_{\mu\nu} \right| } 
- \lambda \sqrt{- g} \right)
+ S_M \left[ g, \Psi_M \right] , 
\label{eq:EiBI_action}
\end{equation}
where $\left| f_{\mu\nu} \right|$ denotes the determinant of $f_{\mu\nu}$ and 
$g \equiv \left| g_{\mu\nu} \right|$. 
$R_{\mu\nu}$ represents the symmetric part of the Ricci tensor and 
is constructed solely from the connection $\Gamma^\alpha_{\beta\gamma}$. 
The matter action $S_M$ is assumed to depend on the metric $g_{\mu\nu}$ 
and the matter field $\Psi_M$ only. 
Furthermore, it can be shown that the action of Eq. (\ref{eq:EiBI_action}) 
is equivalent to the Einstein-Hilbert action when $S_M=0$, and hence the 
theory is identical to GR in vacuum \cite{Ferreira10}. 

The constants $\kappa$ and $\lambda$ are related to the 
cosmological constant by $\Lambda = (\lambda - 1) / \kappa$ such that, 
for small $\kappa R_{\mu\nu}$, the action [Eq. (\ref{eq:EiBI_action})] reduces 
to the Einstein-Hilbert action. We shall fix $\lambda=1$ in this work and 
treat $\kappa$ as the only parameter of the theory. 
The constraints on $\kappa$ based on solar observations \cite{Casanellas12}, 
big bang nucleosynthesis and the existence of neutron stars \cite{Avelino12}
have been studied recently.

In this theory, the spacetime metric $g_{\mu\nu}$ and the connection 
$\Gamma^\alpha_{\beta \gamma}$ are treated as independent fields. 
The field equations are obtained by varying the action [Eq. (\ref{eq:EiBI_action})]
with respect to $g_{\mu\nu}$ and $\Gamma^\alpha_{\beta\gamma}$ seperately, 
and can be written as (with $\lambda = 1$) \cite{Ferreira10}
\begin{eqnarray}
q_{\mu\nu} &=& g_{\mu\nu} + \kappa R_{\mu\nu} , \label{eq:field1} \\
\cr
\sqrt{-q} q^{\mu\nu} &=&  \sqrt{-g} g^{\mu\nu} - 
8\pi \kappa \sqrt{-g} T^{\mu\nu} , \label{eq:field2} 
\end{eqnarray}
where $q_{\mu\nu}$ is an auxiliary metric compatible with the connection
\begin{equation}
\Gamma^\alpha_{\beta\gamma} 
= {1 \over 2} q^{\alpha\sigma} \left( \partial_{\gamma} q_{\sigma\beta}
+\partial_{\beta} q_{\sigma\gamma} - \partial_{\sigma} q_{\beta\gamma} 
\right) ,
\end{equation}
and $q \equiv \left| q_{\mu\nu} \right| $.
The stress-energy tensor $T^{\mu\nu}$ satisfies the same conservation 
equations as in GR:
\begin{equation}
\nabla_\mu T^{\mu\nu} = 0 , 
\label{eq:matter_conserv}
\end{equation}
where the covariant derivative refers to the metric $g_{\mu\nu}$.

\section{Static equilibrium configurations}
\label{sec:Static}

\subsection{Basic equations}

The structure of compact stars in EiBI theory was first studied by 
Pani {\it et al.}~\cite{Pani11,Pani12}. 
Here we present our formulation which uses different 
variables, and hence leads to a different set of differential equations 
which resemble more closely the corresponding equations in GR.  
For a static and spherically symmetric spacetime, the spacetime metric 
$g_{\mu\nu}$ and the auxiliary metric $q_{\mu\nu}$ are taken to be 
\begin{eqnarray}
g_{\mu\nu}dx^\mu dx^\nu &=& -e^{\phi (r)}dt^2+e^{\lambda (r)}dr^2
+f(r)d\Omega^2, \label{eq:metric_bkg}  \\
q_{\mu\nu}dx^\mu dx^\nu &=& -e^{\beta (r)}dt^2+e^{\alpha (r)}dr^2+r^2d\Omega^2.
\end{eqnarray}
Note that the gauge freedom has been used to set $q_{\theta\theta}=r^2$. 
The matter is assumed to be a perfect fluid and is described by the stress-energy tensor
\begin{equation}
T^{\mu\nu} = (\epsilon + P) u^{\mu} u^{\nu} + P g^{\mu \nu} ,
\end{equation}
where $\epsilon$ and $P$ are the energy density and pressure of the fluid,
respectively. The four-velocity of the fluid $u^\mu$ is given by 
$u^\mu = ( e^{-\phi /2}, 0, 0, 0)$ because of the time and spherical symmetry
of the spacetime. The stress-energy tensor is then simplified to 
\begin{equation} 
T^t_t = -\epsilon , \ \ \ 
T^r_r = T^\theta_\theta = T^\varphi_\varphi = P . 
\label{eq:Tmunu_static}
\end{equation}
Note that the indices of $T^{\mu\nu}$ are raised with the spacetime 
metric $g_{\mu\nu}$. 

From Eq.~(\ref{eq:field2}), one can obtain the relations
\begin{equation}
e^{\beta}=e^{\phi}b^3 a^{-1}, \ \  e^{\alpha}=e^{\lambda}a b,
\label{eq:q_static}
\end{equation}
where $a\equiv\sqrt{1+ 8\pi \kappa\epsilon}$ and 
$b\equiv\sqrt{1- 8\pi \kappa P}$.
In addition, $f(r)$ is found to be
\begin{equation}
f(r)=\frac{r^2}{ab} . 
\label{eq:F(r)}
\end{equation}
The field equations of Eq. (\ref{eq:field1}) reduce to the following two 
independent equations 
\begin{eqnarray}
{1\over \kappa}\left( 2+\frac{a}{b^3}-\frac{3}{a
b}\right) &=& \frac{2}{r^2}-\frac{2}{r^2}e^{-\alpha}+\frac{2}{r}
e^{-\alpha}\alpha'  ,   
\label{eq:plus_static}  \\
{1\over \kappa}\left( \frac{1}{ab}+\frac{a}{b^3}-2\right) &=&
-\frac{2}{r^2}+\frac{2}{r^2}e^{-\alpha}+\frac{2}{r}e^{-\alpha}\beta' , 
\label{eq:minus_static}
\end{eqnarray}
where primed quantities denote partial derivatives with respect to $r$.  
The conservation of the stress-energy tensor [Eq.~(\ref{eq:matter_conserv})]
gives 
\begin{equation}
\phi'=-\frac{2P'}{P+\epsilon}\label{eq:conserve_static}.
\end{equation}
Eqs. (\ref{eq:plus_static}) \-- (\ref{eq:conserve_static}) can be combined to 
obtain the following two first-order differential equations:
\begin{eqnarray}
\frac{d P}{d r} &=& - \left[ {1\over 2\kappa} \left( \frac{1}{a
b}+\frac{a}{b^3}-2 \right) r + \frac{2m}{r^2} \right]
\left[1-\frac{2m}{r}\right]^{-1} \nonumber
\\
&&\times\left[\frac{2 }{ \epsilon + P }+\frac{\kappa}{2} 
\left( \frac{3}{b^2} + \frac{1}{a^2 c_s^2} \right) \right]^{-1} , 
\label{eq:P'}
\\
\frac{d m}{d r} &=& {1\over 4 \kappa} \left( 2-\frac{3}{ab}
+\frac{a}{b^3} \right) r^2 , 
\label{eq:m'} 
\end{eqnarray}
where the speed of sound $c_s$ is calculated by $c_s^2 = dP/d\epsilon$. The 
function $m(r)$ is defined by
\begin{equation}
e^{-\lambda}=\left( 1-\frac{2m}{r} \right) ab .
\label{eq:e_lambda}
\end{equation}
With a given EOS $P=P(\epsilon)$, the structure of a 
hydrostatic equilbrium star in EiBI gravity is obtained by solving 
Eqs.~(\ref{eq:P'}) and (\ref{eq:m'}). By expanding $a$ and $b$ in series of 
$\kappa$, it can be shown that these equations reduce to the corresponding 
structure equations in GR when $\kappa \rightarrow 0$ 
(see, e.g., Ref. \cite{Shapiro83}).

\subsection{Boundary conditions and numerical scheme}

To construct a static compact star with a given EOS, we first integrate
Eqs.~(\ref{eq:P'}) and (\ref{eq:m'}) by specifying the central density 
$\epsilon_c$, and hence the central pressure $P_c$, and setting $m(0)=0$. 
The radius of the star $R$ is defined by the condition $P(R)=0$. As discussed
above, EiBI gravity is equivalent to GR in vacuum. Hence, the interior 
solution should match smoothly to the Schwarzschild solution at the stellar
surface. 
It can be checked that the required conditions are 
$\epsilon(R)=P(R)=0$ and hence $a(R)=b(R)=1$. We then have 
$ e^{-\alpha(R)}= e^{-\lambda(R)} = (1 - 2M/R)$, where $M\equiv m(R)$ is the 
mass of the star. 
We also require that $e^{\beta(R)} = e^{ \phi(R)} = (1- 2M/R)$. 
The function $\beta(r)$, and hence $\phi(r)$ because of 
Eq.~(\ref{eq:q_static}), can now be obtained by integrating 
Eq.~(\ref{eq:minus_static}) backward from the surface to the center. 
We have now completed the interior solution of the star. 
It should be noted that the appearance of the terms 
$a=\sqrt{1+8\pi\kappa \epsilon}$ and $b=\sqrt{1-8\pi\kappa P}$ in the theory 
imposes the following conditions (see also Refs. \cite{Pani11,Pani12}):
\begin{eqnarray}
8\pi \kappa P_c &<& 1 \ , \ \ {\rm for }\ \kappa > 0 , 
\label{eq:max_pressure}  \\
8\pi {| \kappa |} \epsilon_c &<& 1 \ , \ \ {\rm for }\ \kappa < 0 .
\label{eq:max_density}
\end{eqnarray}

\section{Radial oscillations}
\label{sec:perturbation}

\subsection{Equations for radial perturbations}

In order to check the stability of compact stars constructed in EiBI gravity,
we need to study the radial oscillation modes of these stars. 
The corresponding study in GR was first performed by Chandrasekhar almost
fifty years ago \cite{Chandra64} (see also Ref. \cite{Kokkotas01} for a more 
recent study). 
Here we shall derive an eigenvalue equation [see Eq.~(\ref{eq:radial_master}) 
below] which allows us to obtain the frequencies of radial oscillation modes 
of compact stars in EiBI gravity. 
Since the calculation is somewhat tedious, we shall thus only outline 
the main steps of the derivation in the following.

We assume that the static background star is perturbed radially so that 
spherical symmetry is maintained. The spacetime metric can still be 
written as Eq.~(\ref{eq:metric_bkg}), but now the components $g_{tt}$ and 
$g_{rr}$ depend on both $r$ and $t$ (see, e.g., Ref. \cite{Weinberg72}):
\begin{equation}
g_{\mu \nu} dx^\mu dx^\nu = -e^{\phi(t,r)}dt^2 + e^{\lambda(t,r)}dr^2
+ f(r) d\Omega^2 . 
\end{equation}
The auxiliary metric $q_{\mu\nu}$ takes the form 
\begin{equation}
q_{\mu\nu}dx^\mu dx^\nu = -e^{\beta(t,r)} dt^2 + e^{\alpha(t,r)}dr^2
+ 2 \eta(t,r) dt dr + r^2 d\Omega^2 , 
\end{equation} 
where the $tr$ component [$q_{tr} \equiv \eta(t,r)$] is in general 
nonzero. 
The four-velocity of the fluid is now given by 
\begin{equation}
u^\mu = (-e^{\phi /2}, e^{-\phi /2} \dot{\xi}, 0, 0) , 
\end{equation}
where $\xi$ is the Lagrangian displacement and 
$\dot\xi \equiv \partial \xi/\partial t$. 
To linear order in $\xi$, the 
nonvanishing components of the stress-energy tensor are 
\begin{eqnarray} 
&& T^t_t = -\epsilon , \ \ \ 
T^r_r = T^\theta_\theta = T^\varphi_\varphi = P , \\
\cr
&& T^r_t = - (\epsilon_0+P_0) \dot\xi , 
\label{eq:T_tr}
\end{eqnarray}
where $\epsilon_0$ and $P_0$ refer to the energy density and pressure of
the unperturbed static background, respectively. It should be noted that 
Eqs.~(\ref{eq:field2}) and (\ref{eq:T_tr}) imply that the auxiliary metric 
function $\eta(t,r)$ is in general nonzero even though $g_{tr} = 0$.

We now consider small radial perturbations on the static background 
solution such that $F(t,r) = F_0(r) + \delta F(t,r)$, where $F$ stands for 
any metric or fluid variable and $F_0$ is its background solution.  
We shall derive the linearized field and matter equations by retaining 
terms only of first order in $\xi$ and $\delta F$. 
To obtain the oscillation mode frequencies, we assume a time dependence 
$e^{i\omega t}$ for all the perturbed quantities. 
Eqs.~(\ref{eq:field1}) and (\ref{eq:field2}) can then be reduced to the 
following equations
\begin{eqnarray}
\delta \alpha &=& - { e^{\alpha} \over r } \chi , \\ 
\chi' &=& - r^2 Q_2 \delta \epsilon , \label{eq:chi_prime} \\ 
\delta\beta &=& \delta\phi - 4\pi \kappa \left[\frac{3\delta P}
{1- 8\pi \kappa P}
+\frac{\delta\epsilon}{1+ 8\pi \kappa\epsilon}\right] ,  \\
\delta\beta' &=& -\frac{e^{\alpha}}{r}\frac{Q_3}{Q_2}\chi'-
\frac{e^{\alpha}}{r}Q_4 \chi , 
\end{eqnarray}
where $\chi \equiv r^2 (\epsilon+P) Q_1 \xi$. In the above, physical 
quantities without ``$\delta$'' are evaluated on the static background. The 
functions $Q_i$ are given explicitly in Appendix~\ref{sec:symbols}. 
On the other hand, the linearized 
conservation equation (\ref{eq:matter_conserv}) becomes
\begin{equation}
e^{\lambda-\phi}(\epsilon + P)\omega^2\chi=\delta
P'+\frac{1}{2}(\epsilon + P)\delta\phi'+\frac{1}{2}(1+c_s^2)\phi'\delta P .
\label{eq:conserve_radial}
\end{equation}
Combining the above linearized equations, we obtain our eigenvalue equation 
for determining the radial oscillation modes
\begin{equation}
\chi'' = - W_1 \chi - W_2 \chi' , 
\label{eq:radial_master}
\end{equation}
where the functions $W_1$ and $W_2$ depend only on the background quantities 
and the frequency square $\omega^2$ (see Appendix~\ref{sec:symbols} for their
expressions).

\subsection{Boundary conditions and numerical scheme}

The Lagrangian displacement $\xi$ must vanish at the center due to 
spherical symmetry. From the definition of $\chi$, this condition is 
equivalent to 
\begin{equation}
\chi(0) = 0 . 
\label{eq:BC_centre}
\end{equation}
At the stellar surface, the appropriate boundary condition is that the 
Lagrangian variation of the pressure vanishes ($\delta P = 0$). 
It can be shown from Eq.~(\ref{eq:chi_prime}), with the conditions
$P(R)=\epsilon(R)=0$, that the boundary condition is equivalent to 
the requirement that the displacement $\xi$ is finite at the surface. From 
the definition of $\chi$, the boundary condition is thus equivalent to 
\begin{equation}
\chi (R)=0 .
\label{eq:BC_surface}
\end{equation} 

To find $\omega^2$ numerically, the shooting method is applied. 
We first decompose Eq.~(\ref{eq:radial_master}) into two first-order 
differential equations for $\chi$ and $\chi'$. 
In practice, we choose a trial eigenvalue $\omega^2$ and start our numerical 
integration at a point near the center. 
Note that the regularity condition of $\chi$ 
implies that $\chi \sim r^3$ for small $r$. We integrate up to the stellar
surface and check whether Eq.~(\ref{eq:BC_surface}) is satisfied. 
The eigenvalue is obtained if the trial $\omega^2$ can satisfy the boundary
condition. Otherwise, the integration is repeated with a different trial 
$\omega^2$.

\section{Equation of state} 
\label{sec:eos} 

In constructing a compact star, one has to specify an EOS model that
gives the relation between $P$ and $\epsilon$.
Although polytropic EOSs are often used in compact star simulations, they are 
oversimplified and cannot reflect the complexity of nuclear matter. 
On the other hand, realistic EOS models are usually presented in tabulated 
forms and one needs in general to perform numerical interpolations in the 
study. 
Alternatively, one can also use piecewise polytropic models to fit many
tabulated EOSs in different density regions \cite{Read09}. 

The above two techniques in general work well for studying compact stars in 
GR. 
However, we found that they are not good enough for studying the oscillation 
modes of compact stars in EiBI gravity. The reason is that the eigenvalue 
equation [Eq.~(\ref{eq:radial_master})] involves the derivative $c_s'$, which 
is proportional to $d^2 P / d \epsilon^2$. 
It is noted that, for the corresponding study in GR, one only 
needs to calculate $dP/d \epsilon$ \cite{Kokkotas01}. 
In order to apply realistic EOS models in our study, we use smooth analytic 
functions to model the tabulated EOS models so that the second derivative
$d^2 P / d \epsilon^2$ can be computed analytically, and hence
numerical errors can be reduced. In particular, we use the following 
analytical representation suggested by Haensel and Potekhin~\cite{Haensel04}: 
\begin{eqnarray}
{\tilde P}
&=&\frac{a_1+a_2{\tilde \epsilon}+a_3{\tilde \epsilon}^3}
{1+a_4{\tilde\epsilon} }f(a_5({\tilde\epsilon}-a_6)) \nonumber
\\
&&+(a_7+a_8{\tilde\epsilon})f(a_9(a_{10}-{\tilde\epsilon})) \nonumber
\\
&&+(a_{11}+a_{12}{\tilde\epsilon})f(a_{13}(a_{14}-{\tilde\epsilon})) \nonumber
\\
&&+(a_{15}+a_{16}{\tilde\epsilon}) f(a_{17}(a_{18}-{\tilde\epsilon})) , 
\label{eq:analytic} 
\end{eqnarray}
where ${\tilde P} = \log(P/$dyn cm$^{-2})$, 
${\tilde \epsilon} = \log(\epsilon /$g cm$^{-3})$, and 
$f(x) = 1/(e^x + 1)$. The 18 constants $a_i$ are fitting 
parameters. 
For a given tabulated EOS, we use the 
Levenberg-Marquardt method~\cite{Press92} to determine the set of 
parameters $a_i$ that best fits the EOS data points.
In this work, we consider the following eight realistic EOSs: 
model A~\cite{Pandharipande71}, model APR~\cite{Akmal98}, 
model BBB2~\cite{Baldo97}, model C~\cite{Bethe74}, 
model FPS~\cite{Lorenz93}, model SLy4~\cite{Douchin00}, 
model UU~\cite{Fiks88,Negele73}, and model WS~\cite{Fiks88,Lorenz93}.

\begin{figure}
\centering
\includegraphics*[width=7.5cm]{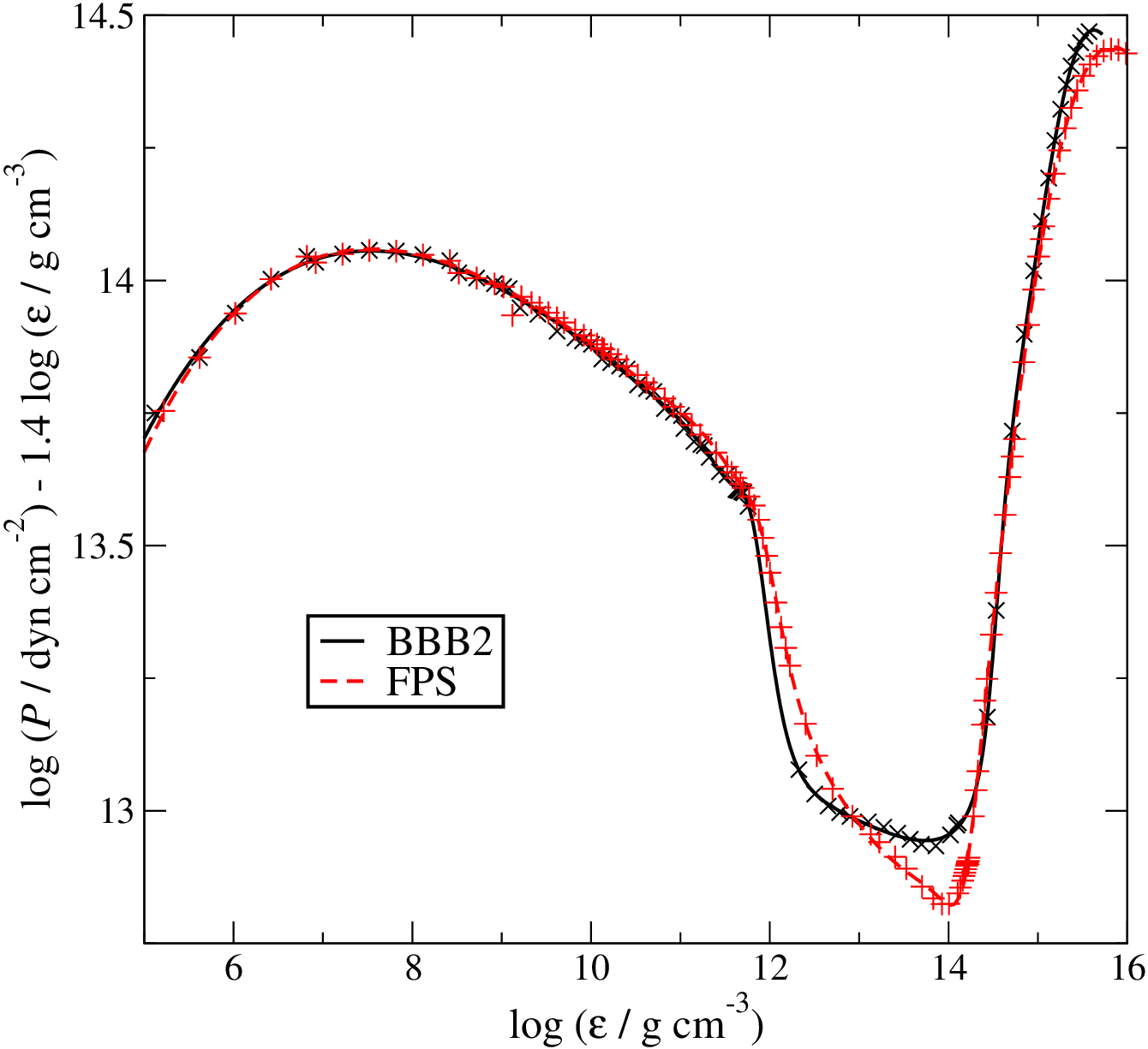}
\caption{Plot of the analytic fit of the BBB2 and FPS models. The
crosses and pluses represent the data points in the EOS tables and
the lines are the analytic fit functions 
(see also Fig.~2 of Ref. \cite{Haensel04}).  } 
\label{fig:analytic_fit}
\end{figure}

To show the accuracy of our analytical representations of the tabulated EOS, 
we plot in Fig.~\ref{fig:analytic_fit} the analytical fits to the BBB2 
(solid line) and FPS (dashed line) EOS models. 
The original EOS data points are denoted by the cross (BBB2) and plus (FPS) 
symbols in the figure. It is seen from the figure that the analytic fits match 
the data points of the tabulated EOSs very well. 
Furthermore, we find that the standard deviations between the 
original EOS data and fitting data points are in general of the order 
$10^{-2}$ or less for all the EOS models.

\section{Results}
\label{sec:results}

\begin{figure*}
  \begin{minipage}{0.5\linewidth}
  \centering
  \includegraphics*[width=7.5cm]{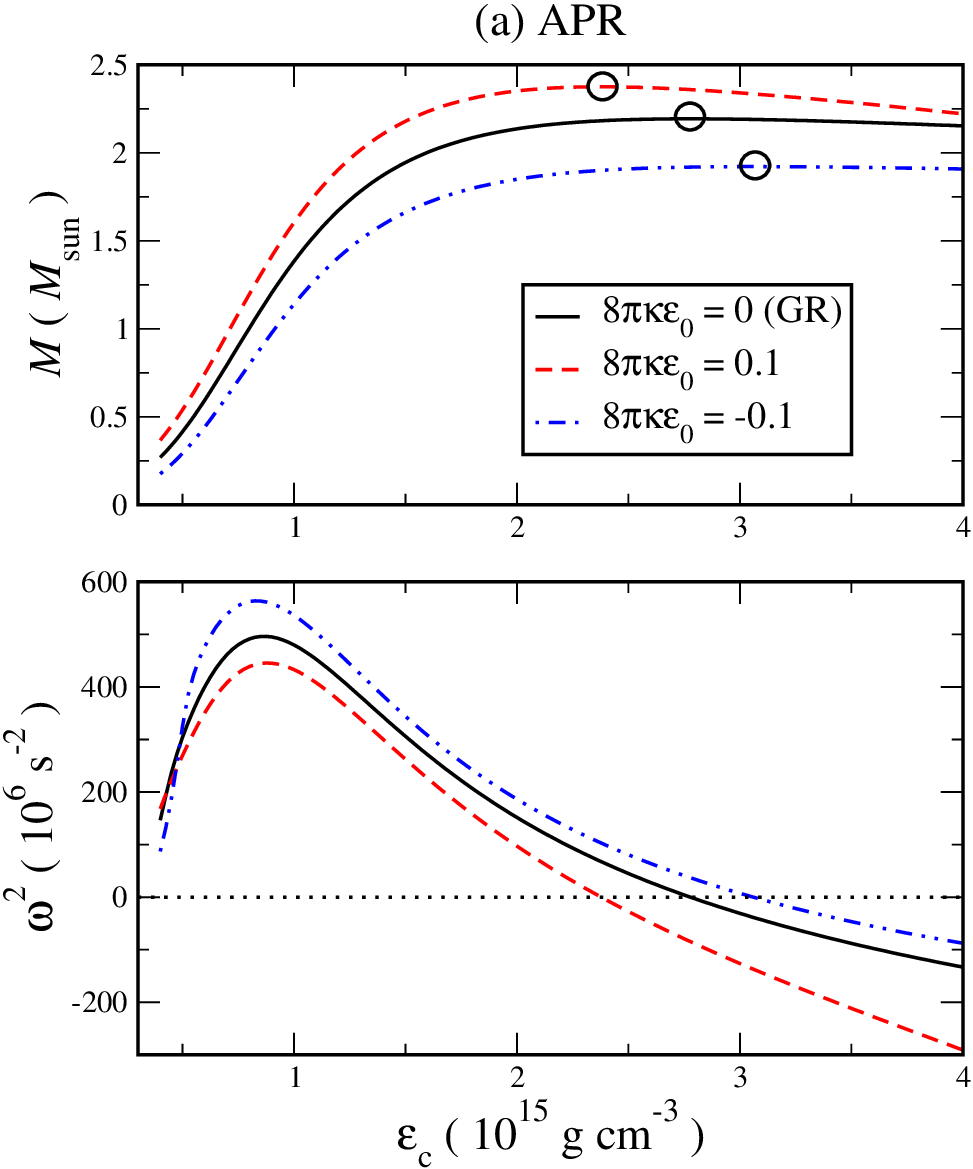}
  \end{minipage}%
  \begin{minipage}{0.5\linewidth}
  \includegraphics*[width=7.5cm]{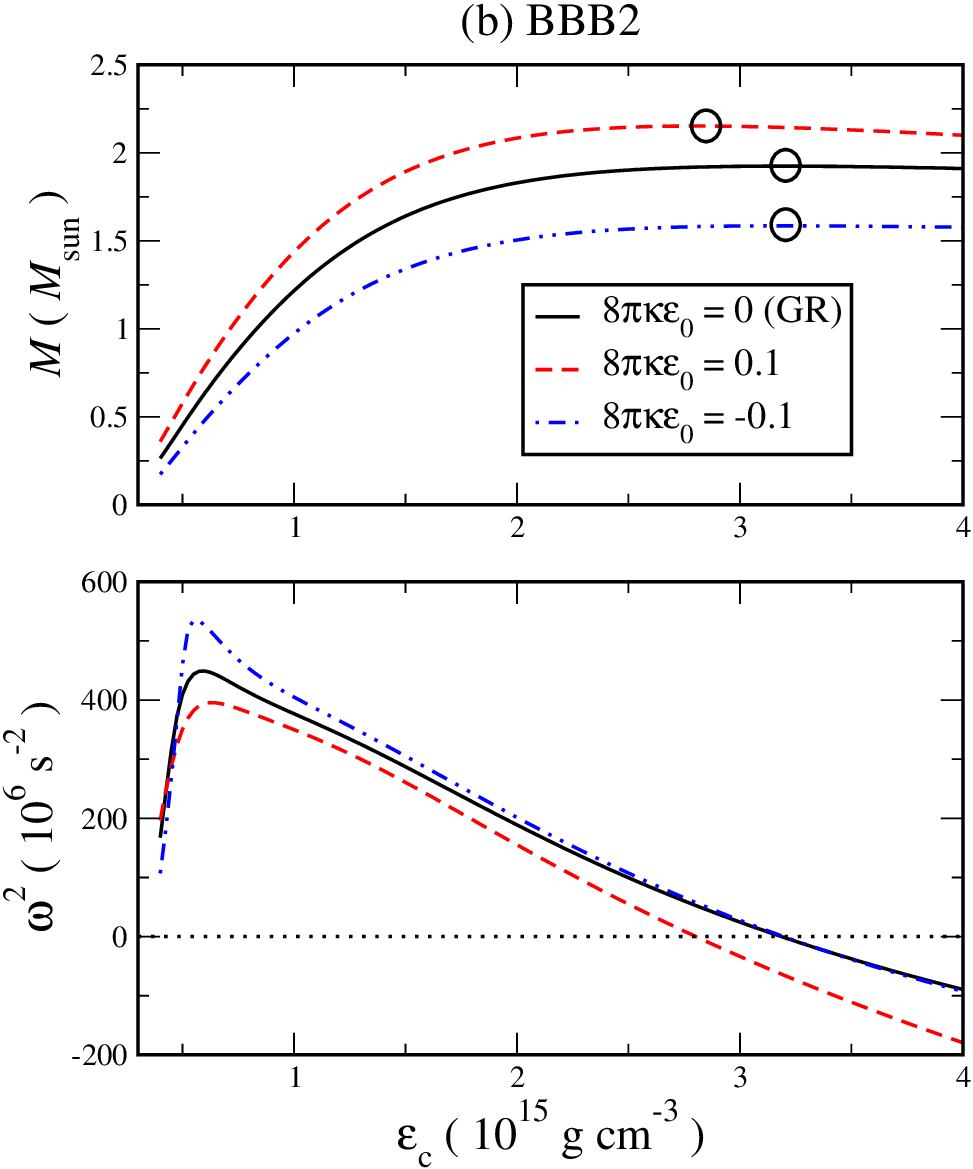}
 \end{minipage}%
 \\
 ~~\\
 \begin{minipage}{0.5\linewidth}
  \centering
  \includegraphics*[width=7.5cm]{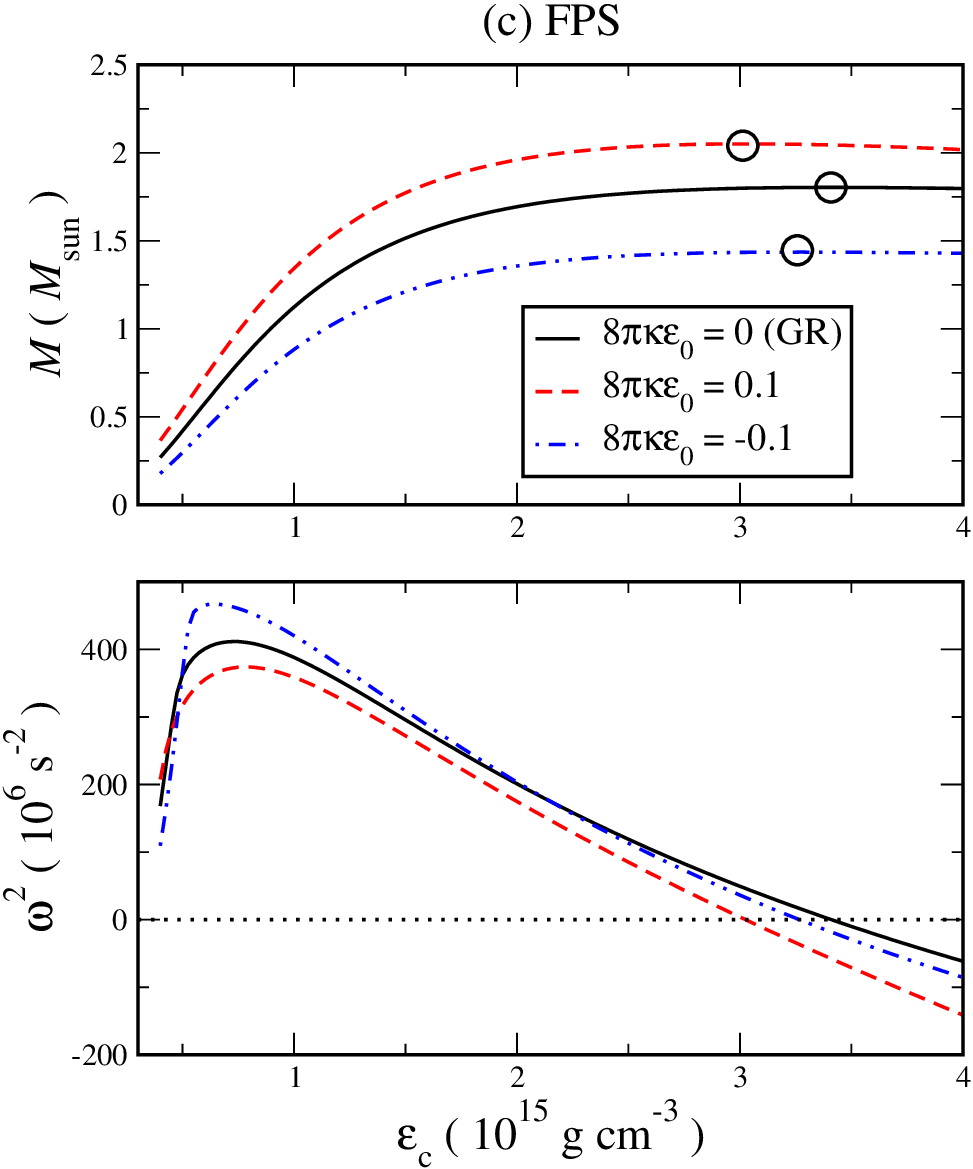}
  \end{minipage}%
  \begin{minipage}{0.5\linewidth}
  \includegraphics*[width=7.5cm]{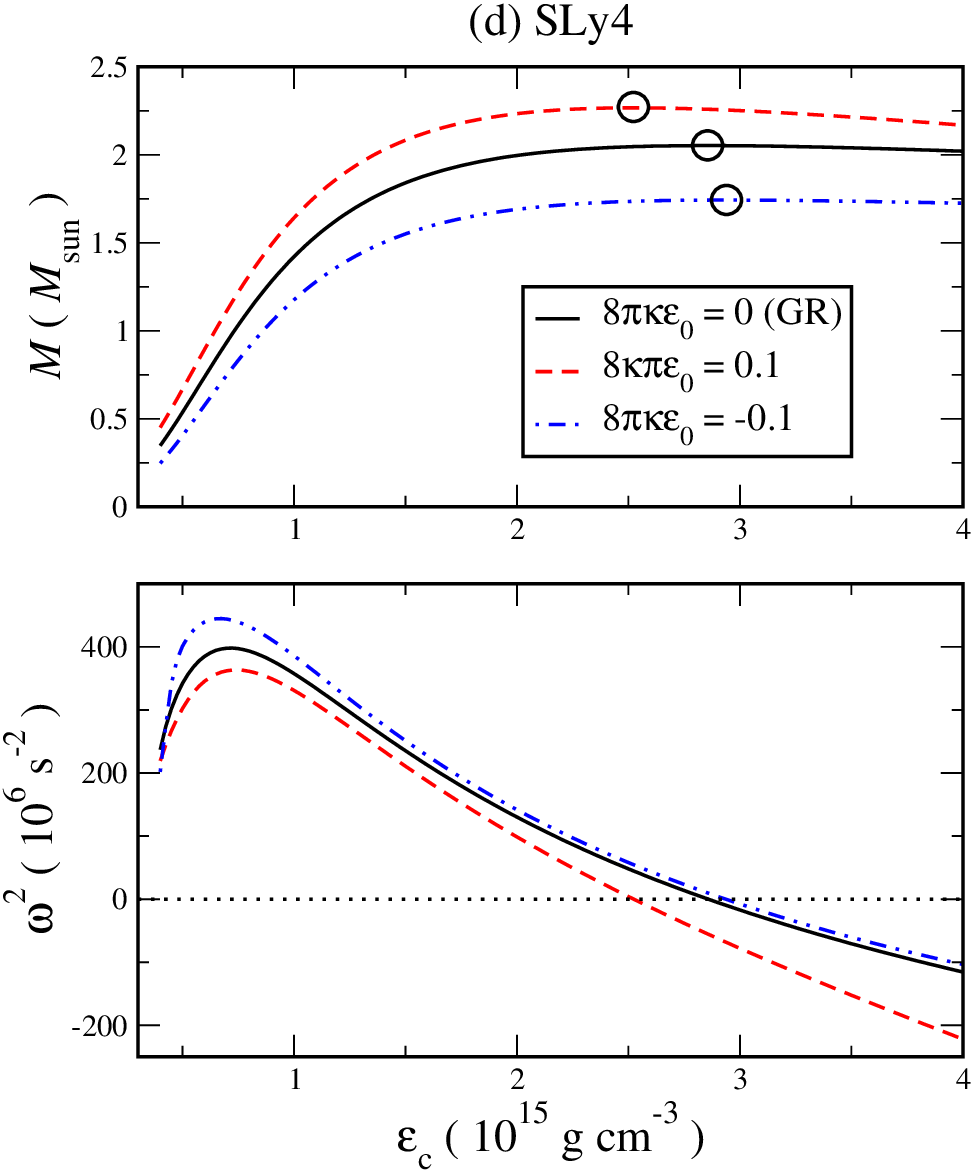}
  \end{minipage}%
  \caption{ Gravitational mass $M$ and fundamental mode frequency square 
$\omega^2$ plotted against the central density $\epsilon_c$ for the 
four EOS models: (a) APR, (b) BBB2, (c) FPS and (d) SLy4. Three different 
values of $\kappa$ are considered. The circle on each $M$-$\epsilon_c$ 
curve corresponds to the maximum-mass configuration. }
  \label{fig:mass_omega}
\end{figure*}

In this section, we shall construct static equilbrium compact stars and 
study their stability in EiBI gravity. As discussed in Sec.~\ref{sec:EiBI}, 
the free parameter of the theory is $\kappa$. For $\kappa = 0$, the theory is 
equivalent to GR. On the other hand, the more interesting case is 
$\kappa > 0$ because it is the regime where novel properties of EiBI 
gravity exist. In this regime, the theory leads to a nonsingular 
cosmological model \cite{Ferreira10} and the existence of pressureless 
stars \cite{Pani11,Pani12}. 
In this work, we shall consider three different values of $\kappa$ 
defined by $8\pi\kappa\epsilon_0 = -0.1,\ 0,\ 0.1$, 
where $\epsilon_0 = 10^{15} {\rm g \ cm}^{-3}$. 
These values of $\kappa$ are consistent with the recent constraint set by
the existence of neutron stars as proposed in Ref. \cite{Avelino12}, though it 
should be noted that we use units where $G=c=1$ in this work.

\begin{figure}
\centering
\includegraphics*[width=6.5cm]{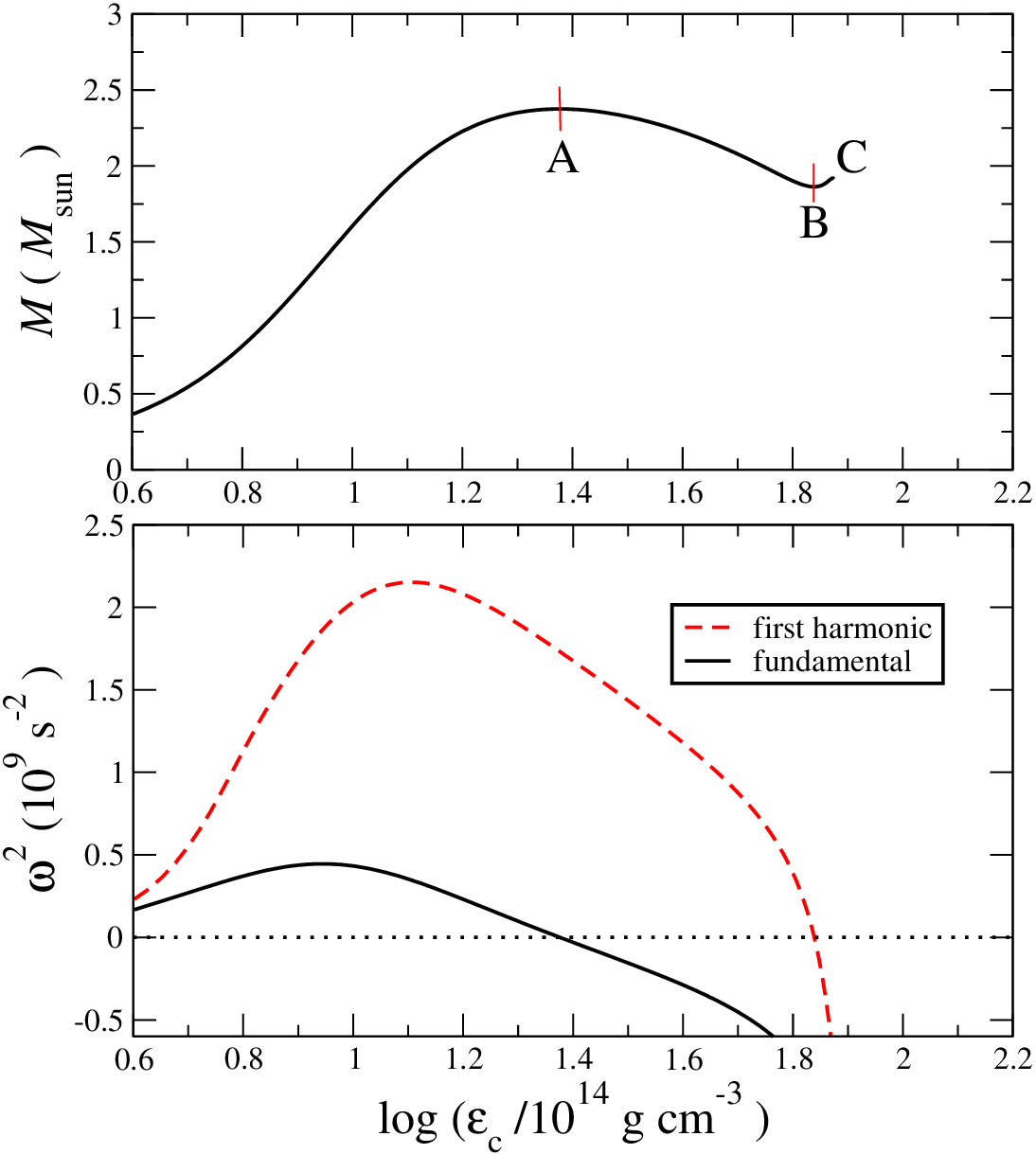}
\caption{ Upper panel: Gravitational mass as a function of the central 
density for compact stars modeled by the APR EOS and 
$8\pi\kappa\epsilon_0 = 0.1$.
The curve terminates 
at the point C where Eq.~(\ref{eq:max_pressure}) is violated. 
Lower panel: Frequency square of the fundamental (solid line) and 
first harmonic (dashed line) modes as a function of the central density. } 
\label{fig:first_harmonic_apr}
\end{figure}

To demonstrate the stability of compact stars in EiBI gravity, we show
in Figs.~\ref{fig:mass_omega} (a)\--2(d) the numerical results for the 
APR, BBB2, FPS, and SLy4 EOS models. 
For each case, we plot the gravitational mass $M$ and the frequency square 
$\omega^2$ of the fundamental oscillation mode as a function of the central 
density $\epsilon_c$ in the upper and lower panels, respectively. 
The circle on each $M$-$\epsilon_c$ curve corresponds to the maximum-
mass stellar configuration. 
It is seen from the figures that the $M$-$\epsilon_c$ relation in EiBI 
gravity is qualitatively similar to that in GR. The stellar mass increases 
with the central density until it reaches a maximum. In GR, it is known that
the mode frequency passes through zero at the central density corresponding 
to the maximum-mass configuration. This critical density corresponds to the
onset of dynamical instability in the sense that stellar models beyond this 
critical point are unstable against radial perturbations. 
Our results show that this property is also true in EiBI gravity. 
We see that $\omega^2$ also passes through zero at the maximum-mass 
configuration in EiBI gravity. 
For a given EOS, stellar models with central densities less than the critical 
density are stable becuase the fundamental mode frequency square 
$\omega^2 > 0$. We have thus demonstrated the stability of compact stars in 
EiBI gravity.

For a given EOS model, we see that a negative value of $\kappa$ in general 
decreases the maximum mass of a neutron star compared to the case of GR. 
On the other hand, for a positive value of $\kappa$, EiBI gravity can lead 
to a much larger maximum mass. 
As pointed out in Ref. \cite{Pani12}, this has the interesting implication that 
some softer EOS models, which are ruled out in GR by the recent discovery of a 
neutron star with $M\approx 1.97 M_\odot$~\cite{Demorest10}, would be revived 
in EiBI gravity. 
For example, the maximum mass of a neutron star for the FPS EOS in GR 
is $M=1.8 M_\odot$. However, it can increase to $2 M_\odot$ in EiBI 
gravity for the case $8\pi\kappa\epsilon_0=0.1$.  
In Table~\ref{tab:max_mass}, we list the mass and radius of the maximum-mass 
configuration for each EOS model we consider in this work.  
Furthermore, we list the mass and fundamental mode frequency as a function
of the central density for each EOS model in Appendix~\ref{sec:tables}. 

\begin{table}
\caption{Mass and radius of the maximum-mass stellar configuration for 
each EOS model. 
Mass $M_{\rm max}$ is expressed in solar mass units, and radius $R$ is expressed 
in km.}
\begin{center}
\begin{tabular}{|l |l |l |l |l |l |l|}
\hline
 &\multicolumn{2}{c|}{$8\pi\kappa\epsilon_0=-0.1$}&\multicolumn{2}{c|}{$8\pi\kappa\epsilon_0=0$}&\multicolumn{2}{c|}{$8\pi\kappa\epsilon_0=0.1$}\\
\hline
\multicolumn{1}{|c|}{EOS}&\multicolumn{1}{|c|}{$M_{\textrm{max}}$}&\multicolumn{1}{|c|}{$R$}&\multicolumn{1}{|c|}{$M_{\textrm{max}}$}&\multicolumn{1}{|c|}{$R$}&\multicolumn{1}{|c|}{$M_{\textrm{max}}$}&\multicolumn{1}{|c|}{$R$}\\
\hline\hline A &  1.240 & 7.65&1.656 & 8.21& 1.916& 8.98\\
APR &1.922&9.23 & 2.191& 9.82& 2.375& 10.47\\
BBB2& 1.585 &8.83& 1.922& 9.36& 2.152& 10.03\\
C &1.488& 9.29 &1.838 &9.68& 2.089 &10.33 \\
FPS &1.434 &8.60 &1.800 &9.10& 2.050 &9.79 \\
SLy4 &1.744 &9.33 &2.052 &9.86 &2.268& 10.51 \\
UU &1.932 &9.08 &2.196& 9.67 &2.371 &10.34 \\
WS &1.495 &8.90 &1.845 &9.42 &2.094 &10.05\\
\hline
\end{tabular}
\end{center}
\label{tab:max_mass}
\end{table}

It is recalled that Eqs.~(\ref{eq:max_pressure}) and (\ref{eq:max_density}) 
must be fulfilled in order to construct compact stars in EiBI gravity. 
In the upper panel of Fig.~\ref{fig:first_harmonic_apr} we plot $M$ against 
$\epsilon_c$ for compact stars modeled by the APR EOS and 
$8\pi \kappa\epsilon_0=0.1$ again, but now the central density is extended 
to the point C where Eq.~(\ref{eq:max_pressure}) is violated. In the lower 
panel, we plot the frequency square $\omega^2$ of the fundamental and first 
harmonic modes against $\epsilon_c$. 
As we have seen in Fig.~\ref{fig:mass_omega} (a), the fundamental mode becomes 
unstable at the first critical point A. However, we now also see that there 
exists a second critical point B at higher densities. 
Figure~\ref{fig:first_harmonic_apr} shows clearly that it is the first harmonic 
that changes stability at this point. 
The stellar models from B to C are still unstable even though 
$dM/d\epsilon_c > 0$ in this range. Similar to standard neutron stars in GR 
\cite{Shapiro83}, we have thus seen that the criterion $dM/d\epsilon_c > 0 $ 
also does not guarantee stellar stability in EiBI gravity.

\begin{figure}
\centering
\includegraphics*[width=6.5cm]{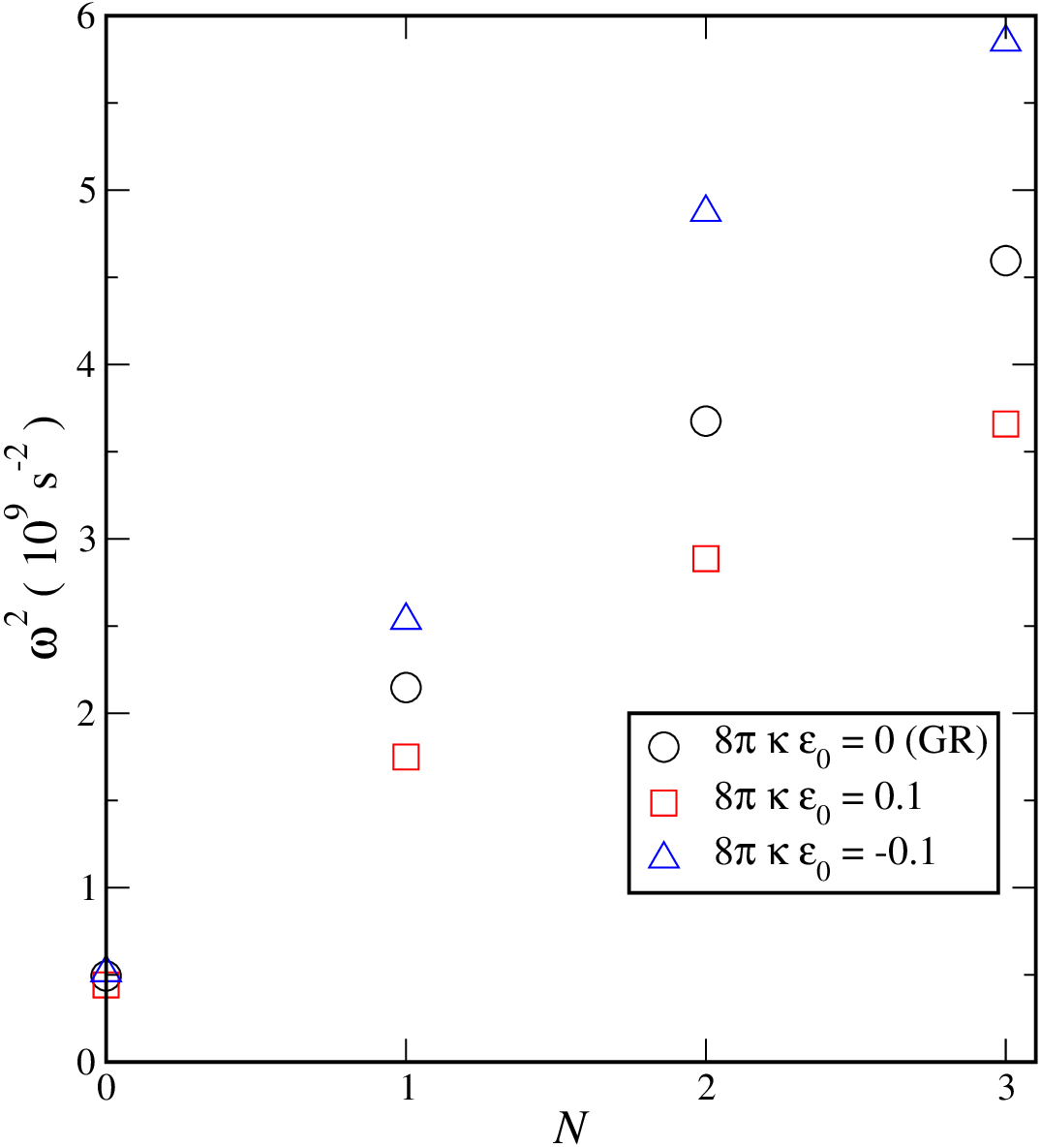}
\caption{Frequency square $\omega^2$ plotted against the number of 
nodes $N$ in the mode eigenfunctions for compact stars modeled by
the APR EOS. The stellar mass $M=1.25 M_\odot$ is fixed. }
\label{fig:mode_apr}
\end{figure}

To end this section, we study how the value of $\kappa$ would affect the 
oscillation mode frequencies. Since the gravitational mass of a compact 
star can usually be measured more accurately than its radius (see 
Ref. \cite{Lattimer07} for a review on neutron star observations), we shall thus 
consider the oscillation modes of compact star models with a given 
mass for different values of $\kappa$. In Fig.~\ref{fig:mode_apr}, we plot the 
frequency square $\omega^2$ against the number of nodes $N$ in the mode 
eigenfunctions. 
The stellar models have the same mass $M=1.25 M_\odot$ and are described by the same APR EOS. Three different cases $8\pi\kappa\epsilon_0=-0.1,\ 0,\ 0.1$ 
are considered as before. Note that the fundamental mode has zero nodes, and 
we see from the figure that the frequency of this mode is insensitive to the 
value of $\kappa$. 
On the contrary, higher-order modes (i.e., modes with larger values of $N$) 
depend more sensitively on the value of $\kappa$. 
In particular, a positive (negative) value of $\kappa$ would decrease 
(increase) the frequencies of higher-order modes. While we only show the 
results of three different values of $\kappa$ in Fig.~\ref{fig:mode_apr}, we 
have seen that this property is in general true for other values.

Finally, we note that the sensitivity of the frequencies of higher-order
modes on $\kappa$ may be understood heuristically by noting that, in the 
nonrelativistic limit of EiBI gravity, the standard Poisson equation in 
Newtonian theory is modified by having an extra source term of the form 
$\kappa \nabla^2 \rho$, where $\rho$ is the mass density~\cite{Ferreira10}.  
A larger variation of density perturbation, which is the case for higher-order modes, would then lead to a larger difference in the stellar dynamics 
comparing to the Newtonian results. 
On the other hand, the fundamental mode eigenfunction has no node and is 
monotonically increasing in the star. Hence, the magnitude of 
$\nabla^2 \delta \rho$ (with $\delta\rho$ being the density perturbation 
associated to the mode) should in general be smaller than 
those of higher-order modes. This makes the frequency of the fundamental mode
not so sensitive to $\kappa$. 
We believe that this property is also true in our fully relativistic study, 
as shown in Fig.~\ref{fig:mode_apr}.

\section{Conclusion}
\label{sec:conclusion}

In this paper, we have studied the equilibrium structure of compact stars in 
EiBI gravity using a set of eight realistic EOS models. 
Our formulation of the structure equations is different from the recent 
works of Pani {\it et al.}~\cite{Pani11,Pani12} in such a way that the 
resulting differential equations resemble more closely the corresponding 
equations in GR. 
We solved the structure equations numerically and found that the maximum mass 
of a neutron star can be larger than that in GR when the parameter $\kappa$ in 
EiBI gravity is positive. 
This implies that softer EOS models, which are ruled out in GR by the recent 
discovery of a neutron star with mass nearly $2 M_\odot$, would be revived in 
EiBI gravity~\cite{Pani12}. 

We have also developed a theory of radial perturbation and studied the 
stability of compact stars in EiBI gravity by calculating the oscillation
mode frequency square $\omega^2$ for the first time. 
In contrast to the situation in GR, since the oscillation equation in EiBI 
gravity involves the second derivative $dP^2/ d\epsilon^2$, we found that 
using standard techniques such as numerical interpolation or piecewise 
polytropic representation \cite{Read09} to handle realistic EOS tables in the 
calculation would not produce reliable numerical results. This is because 
the data points in standard EOS tables are usually not dense enough to allow
an accurate calculation of $dP^2/ d\epsilon^2$.
We thus followed Ref. \cite{Haensel04} and used an 18-parameter analytic 
representation to model the EOS in our calculation. 
It should, however, be emphasized that the analytic fitting is not 
fundamentally essential to our stability analysis. It is employed in this work 
in order to improve numerical accuracy only.

We find that the standard results of stellar stability still hold in EiBI 
gravity. 
For a sequence of stars modeled by the same EOS, we found that the fundamental
mode frequency square passes through zero at the central density
corresponding to the maximum-mass configuration. Similar to the analysis of 
compact stars in GR, this point marks the onset of instability. Stellar models
with central densities less than the critical point are stable because 
$\omega^2 > 0 $. 
Furthermore, we also found that the criterion $dM/d\epsilon_c > 0 $ does not 
guarantee stellar stability.

We have also studied the effects of the parameter $\kappa$ on the oscillation
modes. For a fixed stellar mass, we found that the fundamental mode frequency 
is insensitive to the value of $\kappa$. On the contrary, the 
frequencies of higher-order modes depend strongly on $\kappa$. 
In particular, a positive (negative) value of $\kappa$ would decrease 
(increase) the frequencies of higher-order modes. 
Our results thus suggest that the detection of higher-order radial 
oscillation modes might provide useful constraints on the value of $\kappa$
if they could be excited to large amplitudes. 
Of course, in reality oscillation modes could in general be excited to 
large amplitudes only in some catastrophic situations such as core
collapse supernovae. However, those catastrophic events are in general 
highly nonspherical, and hence nonradial oscillation modes could become more 
relevant. Studying nonradial oscillations of compact stars and the 
corresponding gravitational wave signals would be a natural extension
of our present work. We hope to return to this issue in the future and study 
whether the gravitational wave signals emitted by compact stars could be 
used to constrain EiBI gravity.

\begin{appendix}

\section{List of functions}
\label{sec:symbols}
Here we present the expressions for the various functions $Q_i$ and $W_i$ 
used in Sec.~\ref{sec:perturbation}: 

\begin{eqnarray}
Q_{1}&=&\frac{8\pi}{a^2}\left[1  +  \kappa
e^{-\alpha} \left(\frac{\beta'}{r}-\frac{\alpha'\beta'}{4}+\frac{\beta'^2}{4}
 + \frac{\beta''}{2} \right) \right] \nonumber \\ 
\cr
Q_2 &=& 6\pi\left[\frac{a^2+3b^2}{3a^3b^3}+\frac{a}{b^3}\left(\frac{1}{b^2}-\frac{1}{a^2}\right)c_s^2\right]  , \\
\cr
Q_3 &=& 2\pi\left[\frac{1}{a^2}\left(\frac{a}{b^3}-\frac{1}{ab}\right)
+\frac{c_s^2}{b^2}\left(\frac{3a}{b^3}+\frac{1}{ab}\right)\right] , \\
Q_4&=&\frac{1}{r}+\beta'+ 16\pi e^{\alpha-\beta}\kappa\frac{P+\epsilon}{a^2}\frac{\omega^2}{Q_1} ,  \\
\cr
Q_5&=& 4 \pi \kappa \left[\frac{3 c_s^2}{b^2}+\frac{1}{a^2}\right] , \\
\cr
Q_6 &=&
4\pi \kappa  \left(\frac{6c_s}{b^2}c_s'-\frac{6c_s^2}{b^3}b'-\frac{2}{a^3}a'\right) , \\
\cr
Q_7 &=& r e^{\alpha}Q_3+Q_6-Q_5\left(\frac{2}{r}+\frac{Q_2'}{Q_2}\right) , \\
\cr
W_1&=&Q_2\left[\frac{e^{\lambda-\phi}\omega^2}{Q_1}+\frac{r}{2}(P+\epsilon)
e^{\alpha}Q_4 \right] \times \nonumber \\
&& \left[c_s^2+\frac{1}{2}(P+\epsilon)Q_5\right]^{-1} , \\
\cr
W_2&=&\left[c_s^2+\frac{1}{2}(P+\epsilon)Q_5\right]^{-1} 
\left[2 c_s c_s'+\frac{1}{2}(1+c_s^2)\phi' \right. \nonumber \\ 
&& \left. -c_s^2\left(\frac{2}{r}+\frac{Q_2'}{Q_2}\right)+\frac{1}{2}(P+\epsilon)Q_7\right]   . 
\end{eqnarray}

\section{Numerical data}
\label{sec:tables}

In this appendix, we provide the numerical data for the gravitational mass
$M$ and fundamental mode frequency square $\omega^2$. 
In Tables~\ref{tab:App_first} to \ref{tab:App_last}, we list $M$ 
and $\omega^2$ as a function of the central density $\epsilon_c$ for three
different parameters: $8\pi\kappa\epsilon_0=-0.1$, 0, 0.1. 
In the GR limit ($\kappa=0$), we have checked that stellar 
models obtained by using our analytical representations of 
the EOS and those by the original EOS data (with numerical 
interpolation) agree very well. 
For example, the central densities of the maximum-mass 
configurations obtained by the two methods are in general 
different by about 1\% only.

\begin{table}[h!]
\caption{Model A.}
\begin{center}
\begin{tabular}{|c |c |c |c |c |c |c|}
\hline
 &\multicolumn{2}{|c|}{$8\pi\kappa\epsilon_0=-0.1$}&\multicolumn{2}{c|}{$8\pi\kappa\epsilon_0=0$}&\multicolumn{2}{c|}{$8\pi\kappa\epsilon_0=0.1$}\\
\cline{1-7}
$\epsilon_c$&$M$&$\omega^2$&$M$&$\omega^2$&$M$&$\omega^2$\\
\hline
$10^{15}$ g cm$^{-3}$&$M_{\odot}$&10$^6$~s$^{-2}$&$M_{\odot}$&10$^6$~s$^{-2}$&$M_{\odot}$&10$^6$~s$^{-2}$\\
\hline\hline
3.30 &1.234&92.8 &1.639 &112.5 &1.913&42.9 \\
3.40& 1.236 &76.3&1.644& 98.2&1.915& 24.8 \\
3.50 &1.238 &60.2 &1.647& 84.3& 1.916& 7.0 \\
3.60 &1.239 &44.4 &1.650 &70.8& 1.916 &-10.7 \\
3.70 &1.240 &29.1 &1.652 &57.5& 1.915 &-28.1 \\
3.80 &1.240 &14.1 &1.654 &44.7& 1.913 &-45.5 \\
3.90 &1.240 &-0.5 &1.655 &32.1& 1.911 &-62.6 \\
4.00 &1.240 &-14.7 &1.656 &19.9&1.909 &-79.6 \\
4.10 &1.240 &-28.5 &1.656 &7.9 &1.906 &-96.5 \\
4.20 &1.239& -42.0& 1.656 &-3.8& 1.902 &-113.3\\
\hline
\end{tabular}
\end{center}
\label{tab:App_first}
\end{table}

\begin{table}
\caption{Model APR.}
\begin{center}
\begin{tabular}{|c |c |c |c |c |c |c|}
\hline
 &\multicolumn{2}{|c|}{$8\pi\kappa\epsilon_0=-0.1$}&\multicolumn{2}{c|}{$8\pi\kappa\epsilon_0=0$}&\multicolumn{2}{c|}{$8\pi\kappa\epsilon_0=0.1$}\\
\cline{1-7}
$\epsilon_c$&$M$&$\omega^2$&$M$&$\omega^2$&$M$&$\omega^2$\\
\hline
$10^{15}$ g cm$^{-3}$&$M_{\odot}$&10$^6$~s$^{-2}$&$M_{\odot}$&10$^6$~s$^{-2}$&$M_{\odot}$&10$^6$~s$^{-2}$\\
\hline\hline
2.00 &1.850& 186.4& 2.134 &152.0 & 2.351& 97.1\\
2.20& 1.882 &139.1 &2.165 &104.6 & 2.371 &43.2 \\
2.40 &1.902 &98.8 &2.182 &63.8 &2.375 &-5.0 \\
2.50 &1.909 &80.8 &2.187 &45.5 &2.373 &-27.3 \\
2.60 &1.914 &64.0 &2.190 &28.4  &2.370 &-48.7 \\
2.70 &1.917 &48.4 &2.191 &12.3 &2.364 &-69.2 \\
2.80 &1.919 &33.8 &2.192 &-2.7 &2.358 &-88.8 \\
2.90 &1.921 &20.2 &2.191 &-16.9 & 2.350&-107.8 \\
3.00 &1.922 &7.4 &2.189 &-30.2 &2.341 &-126.2 \\
3.10& 1.922 &-4.7& 2.187 &-42.8 &2.331& -144.1\\
\hline
\end{tabular}
\end{center}
\end{table}

\begin{table}
\caption{Model BBB2.}
\begin{center}
\begin{tabular}{|c |c |c |c |c |c |c|}
\hline
 &\multicolumn{2}{|c|}{$8\pi\kappa\epsilon_0=-0.1$}&\multicolumn{2}{c|}{$8\pi\kappa\epsilon_0=0$}&\multicolumn{2}{c|}{$8\pi\kappa\epsilon_0=0.1$}\\
\cline{1-7}
$\epsilon_c$&$M$&$\omega^2$&$M$&$\omega^2$&$M$&$\omega^2$\\
\hline
$10^{15}$ g cm$^{-3}$&$M_{\odot}$&10$^6$~s$^{-2}$&$M_{\odot}$&10$^6$~s$^{-2}$&$M_{\odot}$&10$^6$~s$^{-2}$\\
\hline\hline 
2.30& 1.550&143.4&1.880 &133.9& 2.130&93.8 \\
2.40& 1.559 &125.1 &1.891&116.6& 2.139&74.1 \\
2.50& 1.567 &107.4& 1.900&99.8&2.145&55.0 \\
2.60& 1.573 &90.3 &1.907 &83.7 & 2.149& 36.3 \\
2.70& 1.577 &73.8 &1.913 &68.1 &2.152 &18.1 \\
2.80 &1.580 &57.8 &1.917 &53.2& 2.152 &0.4 \\
2.90 &1.583 &42.4 &1.919 &38.7 &2.152 &-16.8 \\
3.00 &1.584 &27.6 &1.921 &24.8 & 2.150 &-33.5 \\
3.10 &1.585 &13.3 &1.922 &11.5 &2.147 &-49.8 \\
3.20& 1.585 &-0.5 &1.922 &-1.4 &2.144 &-65.7\\
\hline
\end{tabular}
\end{center}
\end{table}

\begin{table}
\caption{Model C.}
\begin{center}
\begin{tabular}{|c |c |c |c |c |c |c|}
\hline
 &\multicolumn{2}{|c|}{$8\pi\kappa\epsilon_0=-0.1$}&\multicolumn{2}{c|}{$8\pi\kappa\epsilon_0=0$}&\multicolumn{2}{c|}{$8\pi\kappa\epsilon_0=0.1$}\\
\cline{1-7}
$\epsilon_c$&$M$&$\omega^2$&$M$&$\omega^2$&$M$&$\omega^2$\\
\hline
$10^{15}$ g cm$^{-3}$&$M_{\odot}$&10$^6$~s$^{-2}$&$M_{\odot}$&10$^6$~s$^{-2}$&$M_{\odot}$&10$^6$~s$^{-2}$\\
\hline\hline
2.30 &1.471& 77.8 &1.807 &89.5&2.069&70.4 \\
2.40& 1.477 &65.2& 1.816 &77.9& 2.077&57.0\\
2.50 &1.481 &52.8& 1.822& 66.6& 2.082& 43.7\\
2.60& 1.484 &40.6& 1.828 &55.5& 2.086& 30.4\\
2.70& 1.486 &28.7&1.832 &44.6& 2.088 &17.2\\
2.80 &1.487& 17.1 &1.835&33.9 &2.089 &4.1 \\
2.90 &1.488 &5.7 &1.837& 23.4& 2.089 &-9.0 \\
3.00 &1.488 &-5.5 &1.838 &13.2 &2.088& -22.0 \\
3.10 &1.488 &-16.5 &1.838& 3.1 &2.085 &-35.0 \\
3.20 &1.487 &-27.2 &1.838 &-6.7 &2.083& -48.0\\
\hline
\end{tabular}
\end{center}
\end{table}

\begin{table}
\caption{Model FPS.}
\begin{center}
\begin{tabular}{|c |c |c |c |c |c |c|}
\hline
 &\multicolumn{2}{|c|}{$8\pi\kappa\epsilon_0=-0.1$}&\multicolumn{2}{c|}{$8\pi\kappa\epsilon_0=0$}&\multicolumn{2}{c|}{$8\pi\kappa\epsilon_0=0.1$}\\
\cline{1-7}
$\epsilon_c$&$M$&$\omega^2$&$M$&$\omega^2$&$M$&$\omega^2$\\
\hline
$10^{15}$ g cm$^{-3}$&$M_{\odot}$&10$^6$~s$^{-2}$&$M_{\odot}$&10$^6$~s$^{-2}$&$M_{\odot}$&10$^6$~s$^{-2}$\\
\hline\hline
2.60 &1.420 &97.3 & 1.775 &104.9 &2.039 &68.9\\
2.70 &1.425&81.5&1.782 &90.5&2.044 &52.3 \\
2.80 &1.428 &66.1 &1.787& 76.6 &2.047&36.0 \\
2.90 &1.431 &51.2 &1.792 &63.2 &2.049& 20.1 \\
3.00 &1.433 &36.8 &1.795& 50.1 &2.050 &4.4 \\
3.10& 1.434 &22.8 &1.797& 37.5 &2.050&-11.0 \\
3.20 &1.434 &9.2&1.799& 25.3 &2.048 &-26.2 \\
3.30 &1.435 &-3.9& 1.800& 13.4 &2.047& -41.1\\
3.40 &1.434& -16.6 &1.800 &1.9 &2.044 &-55.9 \\
3.50 &1.434 &-29.0 &1.800 &-9.3 &2.041& -70.4\\
\hline
\end{tabular}
\end{center}
\end{table}

\begin{table}
\caption{Model SLy4.}
\begin{center}
\begin{tabular}{|c |c |c |c |c |c |c|}
\hline
 &\multicolumn{2}{|c|}{$8\pi\kappa\epsilon_0=-0.1$}&\multicolumn{2}{c|}{$8\pi\kappa\epsilon_0=0$}&\multicolumn{2}{c|}{$8\pi\kappa\epsilon_0=0.1$}\\
\cline{1-7}
$\epsilon_c$&$M$&$\omega^2$&$M$&$\omega^2$&$M$&$\omega^2$\\
\hline
$10^{15}$ g cm$^{-3}$&$M_{\odot}$&10$^6$~s$^{-2}$&$M_{\odot}$&10$^6$~s$^{-2}$&$M_{\odot}$&10$^6$~s$^{-2}$\\
\hline\hline 
2.00&1.692& 143.4 &1.996 &131.5& 2.235 &99.5 \\
2.10& 1.706 &124.5 &2.011&113.2& 2.248 &79.1 \\
2.20 &1.717 &106.6 &2.024& 95.8& 2.257& 59.4 \\
2.30 &1.725 &89.7 &2.033 &79.2 &2.263 &40.4\\
2.40 &1.732 &73.5 &2.040 &63.5 &2.267 &22.1 \\
2.50 &1.737 &58.2 &2.045 &48.4 &2.268 &4.2 \\
2.60 &1.740 &43.6 &2.049 &34.1 &2.268 &-13.0 \\
2.70 &1.743 &29.8 &2.051 &20.4 &2.266 &-29.8 \\
2.80 &1.744 &16.6 &2.052 &7.4 &2.262 &-46.1 \\
2.90 &1.744 &4.0 &2.052 &-5.0 &2.258 &-62.0\\
\hline
\end{tabular}
\end{center}
\end{table}

\begin{table}
\caption{Model UU.}
\begin{center}
\begin{tabular}{|c |c |c |c |c |c |c|}
\hline
 &\multicolumn{2}{|c|}{$8\pi\kappa\epsilon_0=-0.1$}&\multicolumn{2}{c|}{$8\pi\kappa\epsilon_0=0$}&\multicolumn{2}{c|}{$8\pi\kappa\epsilon_0=0.1$}\\
\cline{1-7}
$\epsilon_c$&$M$&$\omega^2$&$M$&$\omega^2$&$M$&$\omega^2$\\
\hline
$10^{15}$ g cm$^{-3}$&$M_{\odot}$&10$^6$~s$^{-2}$&$M_{\odot}$&10$^6$~s$^{-2}$&$M_{\odot}$&10$^6$~s$^{-2}$\\
\hline\hline 
2.00&1.843 &215.3& 2.127&173.0&2.343 &110.2 \\
2.20& 1.881&166.5&2.163&122.9 &2.366 &51.8 \\
2.40 &1.905 &124.2& 2.183&79.3&2.372&-0.9 \\
2.60& 1.920 &87.4 &2.193 &41.2&2.367 &-48.5 \\
2.80 &1.928 &54.9& 2.196 &7.8 &2.354 &-91.9 \\
3.00&1.931 &26.2 &2.195 &-21.7& 2.338 &-131.5 \\
3.10 &1.932 &13.0 &2.194 &-35.2 &2.328 &-150.1 \\
3.20& 1.932 &0.5 &2.191 &-47.9 &2.318 &-168.0 \\
3.30 &1.932 &-11.3 &2.189 &-60.0 &2.307 &-185.2 \\
3.40 &1.932 &-22.4 &2.186 &-71.2 &2.296& -201.8\\
\hline
\end{tabular}
\end{center}
\end{table}

\begin{table}
\caption{Model WS.}
\begin{center}
\begin{tabular}{|c |c |c |c |c |c |c|}
\hline
 &\multicolumn{2}{|c|}{$8\pi\kappa\epsilon_0=-0.1$}&\multicolumn{2}{c|}{$8\pi\kappa\epsilon_0=0$}&\multicolumn{2}{c|}{$8\pi\kappa\epsilon_0=0.1$}\\
\cline{1-7}
$\epsilon_c$&$M$&$\omega^2$&$M$&$\omega^2$&$M$&$\omega^2$\\
\hline
$10^{15}$ g cm$^{-3}$&$M_{\odot}$&10$^6$~s$^{-2}$&$M_{\odot}$&10$^6$~s$^{-2}$&$M_{\odot}$&10$^6$~s$^{-2}$\\
\hline\hline 
2.30 &1.475 &113.9& 1.812& 118.9& 2.072 &92.3\\
2.40& 1.481&94.6&1.821&101.6 & 2.080& 74.2 \\
2.50 &1.486 &76.3 &1.828& 85.3&2.086& 56.8 \\
2.60& 1.490 &59.0 &1.834& 69.8 &2.090 &40.2 \\
2.70 &1.492 &42.5 &1.838& 55.1& 2.093 &24.2 \\
2.80 &1.494 &26.8& 1.841& 41.1 &2.094 &8.8 \\
2.90& 1.495 &11.8 &1.843 &27.8& 2.094& -6.1 \\
3.00& 1.495 &-2.5 &1.844 &15.0 &2.093 &-20.6\\
3.10& 1.494 &-16.2 &1.845 &2.8& 2.091 &-34.6 \\
3.20 &1.494 &-29.3 &1.845 &-8.9 &2.089 &-48.2\\
\hline
\end{tabular}
\end{center}
\label{tab:App_last}
\end{table}

\end{appendix}

\clearpage



\begin{thebibliography}{0}
\expandafter\ifx\csname natexlab\endcsname\relax\def\natexlab#1{#1}\fi
\expandafter\ifx\csname bibnamefont\endcsname\relax
  \def\bibnamefont#1{#1}\fi
\expandafter\ifx\csname bibfnamefont\endcsname\relax
  \def\bibfnamefont#1{#1}\fi
\expandafter\ifx\csname citenamefont\endcsname\relax
  \def\citenamefont#1{#1}\fi
\expandafter\ifx\csname url\endcsname\relax
  \def\url#1{\texttt{#1}}\fi
\expandafter\ifx\csname urlprefix\endcsname\relax\def\urlprefix{URL }\fi
\providecommand{\bibinfo}[2]{#2}
\providecommand{\eprint}[2][]{\url{#2}}

\end{thebibliography}


\begin{thebibliography}{}

\bibitem{Will2006}
C.~M. Will, Living Rev. Relativity \textbf{9}, 3 (2006). 

\bibitem{Clifton2012}
T. Clifton, P.~G. Ferreira, A. Padilla, and C. Skordis, 
Phys. Rep. \textbf{513}, 1 (2012).   

\bibitem{Ferreira10}
M. Ba\~{n}ados and P.~G. Ferreira, Phys. Rev. Lett. \textbf{105},
011101 (2010).

\bibitem{Pani11}
P. Pani, V. Cardoso, and T. Delsate, Phys. Rev. Lett. \textbf{107},
031101 (2011).

\bibitem{Pani12}
P. Pani, T. Delsate, and V. Cardoso, Phys. Rev. D \textbf{85}, 084020 (2012).

\bibitem{Casanellas12}
J. Casanellas, P. Pani, I. Lopes, and V. Cardoso, Astrophys. J.
\textbf{745}, 15 (2012).

\bibitem{Avelino12}
P.~P. Avelino, Phys. Rev. D \textbf{85}, 104053 (2012).

\bibitem{Shapiro83}
S.~L. Shapiro and S.~A. Teukolsky, \textit{Black Holes, White Dwarfs, and
Neutron Stars: The Physics of Compact Objects} (John Wiley and Sons, New
York, 1983). 

\bibitem{Chandra64}
S. Chandrasekhar, Astrophys. J. \textbf{140}, 417 (1964).

\bibitem{Kokkotas01}
K. Kokkotas and J. Ruoff, Astron. Astrophys. \textbf{366}, 565 (2001).

\bibitem{Weinberg72}
S. Weinberg, \textit{Gravitation and Cosmology} (John Wiley \& Sons,
New York, 1972).


\bibitem{Read09}
J.~S. Read, B.~D. Lackey, B.~J. Owen, and J.~L. Friedman, Phys. Rev. D
\textbf{79}, 124032 (2009). 


\bibitem{Haensel04}
P. Haensel and A.~Y. Potekhin, Astron. Astrophys. \textbf{428}, 191
(2004).

\bibitem{Press92}
W.~H. Press, S.~A. Teukolsky, W.~T. Vetterling and B.~R. Flannery, 
\textit{Numerical Recipes in FORTRAN: The Art of Scientific Computing}
(Cambridge University Press, Cambridge, 1992). 


\bibitem{Pandharipande71}
V. Pandharipande, Nucl. Phys. A\textbf{174}, 641 (1971).

\bibitem{Akmal98}
A. Akmal, V.~R. Pandharipande, and D.~G. Ravenhall, Phys. Rev. C \textbf{58},
1804 (1998).

\bibitem{Baldo97}
M. Baldo, I. Bombaci, and G.~F. Burgio, Astron. Astrophys. \textbf{328}, 274 
(1997).

\bibitem{Bethe74}
H.~A. Bethe and M. Johnson, Nucl. Phys. A\textbf{230}, 1 (1974).


\bibitem{Lorenz93}
C.~P. Lorenz, D.~G. Ravenhall, and C.~J. Pethick, Phys. Rev. Lett. 
\textbf{70}, 379 (1993).


\bibitem{Douchin00}
F. Douchin and P. Haensel, Phys. Lett. B \textbf{485}, 107 (2000).


\bibitem{Fiks88}
R.~B. Wiringa, V. Fiks, and A. Fabrocini, Phys. Rev. C \textbf{38},
1010 (1988).

\bibitem{Negele73}
J.~W. Negele and D. Vautherin, Nucl. Phys. A\textbf{207}, 298 (1973).



\bibitem{Demorest10}
P. Demorest, T. Pennucci, S. Ransom, M. Roberts and J. Hessels, Nature (London) \textbf{467}, 1081 (2010). 


\bibitem{Lattimer07}
J.~M. Lattimer and M. Prakash, Phys. Rep. \textbf{442}, 109 (2007). 



\end{thebibliography}
\end{document}